\documentclass{elsart}
\usepackage{epsf}

\begin{document}
\begin{frontmatter}

\title{The bulk of the stock market correlation matrix is not pure noise}

\author{J. Kwapie\'n$^{1}$, S. Dro\.zd\.z$^{1,2}$, P. O\'swi\c 
ecimka$^{1}$}

\address{$^1$Institute of Nuclear Physics, Polish Academy of Sciences, 
PL--31-342 Krak\'ow, Poland \\
$^2$Institute of Physics, University of Rzesz\'ow, PL--35-310 Rzesz\'ow,
Poland}

\begin{abstract}

We analyse the structure of the distribution of eigenvalues of the stock
market correlation matrix with increasing length of the time series
representing the price changes. We use 100 highly-capitalized stocks from
the American market and relate result to the corresponding ensemble of
Wishart random matrices. It turns out that systematically more eigenvalues
stay beyond the borders prescribed by this variant of the Random Matrix
Theory (RMT). This may indicate that even the bulk of the spectrum of the
stock market correlation matrix carries some sort of correlations that are
masked by a measurement noise when the time series used to construct the
matrix are short. We also study some other characteristics of the "noisy"
eigensignals, like their return distributions, temporal correlations or 
their multifractal spectra and the results support the above conclusions.

\end{abstract}

\end{frontmatter}

\section{Introduction}

The formalism of correlation matrix is widely used in contemporary finance
both on theoretical and on practical level in order to estimate
correlation structure of a financial market or in order to create
portfolios characterized by given properties. This formalism is especially
important for various risk management techniques like e.g. the Markowitz
optimal portfolio theory~\cite{markowitz52} which helps one constructing
portfolios offering minimal risk at a given return or maximal return at a
given risk. This theory uses eigenvalues of the correlation matrix as a
measure of the portfolio's risk: no matter which eigenvalue one uses, it
is fully informative and a corresponding portfolio is therefore valid.
However, recently such understanding of the portfolio selection has been
strongly challenged in a series of papers linking the correlation matrix
formalism and the portfolio theories with Random Matrix Theory (RMT) and
the ensemble of Wishart matrices (sample random correlation matrices) in
particular~\cite{laloux99,plerou99a}. Accordingly, the current view is
that only a few eigenvalues and eigenvectors of the correlation matrix are
important in practice and carry any significant information about the
market, while all others describe nothing more than pure noise. Thus, only
those few portfolios which correspond to the non-random eigenstates of the
correlation matrix can be in fact regarded as a potential investment
target. This observation profoundly restricting the applicability of
classical portfolio theories rises a question of how to reconcile the two
contradicting views on the RMT-like portfolios. This problem is at present
one of central issues of econophysics and is intensively studied by many
different groups. One of directions in which current analyses go is the
so-called denoising of the correlation matrices which aims at removing the
estimation errors of the correlations due to finite size of empirical
data~\cite{pafka02,pafka03,pafka04,burda04}.  Another possible approach is
to look into the empirical data and to answer the question what is there
really random and what is not. In order to do this, one can firstly
investigate the properties of dynamics of different portfolios calculated
by using the correlation matrix framework, to compare characteristics of
random and non-random ones, and, possibly, to identify those signatures of
the dynamics which represent some non-random phenomena even in the case of
the RMT-like eigenstates. In the present paper we would like to follow
this idea and draw some conclusions based on data from the American stock
market.

\section{Formalism}

In general, a portfolio $P$ consists of a number of securities $X_s,
s=1,...,N$ associated with weights $w_s$ characterizing the fraction of
total amount of capital invested in a particular security. Return of such
a portfolio after time $\Delta t$ is the weighted sum of logarithmic price 
increments $g_s(\Delta t)=\ln p_s(t+\Delta t) - \ln p_s(t)$ of individual 
securities $X_s$:
\begin{equation}
G_P(\Delta t) = \sum_{s=1}^N w_s g_s(\Delta t).
\label{portreturn}
\end{equation}
By fixing the time scale $\Delta t$ and creating a time series of length
$T$ from consecutive discrete-time portfolio's returns
$\{G_P(j)\}_{j=1}^T$ it is also possible to investigate dynamics of the
portfolio in time. Although every possible set of weights define certain
portfolio, from an investor's perspective only those portfolios which 
are characterized by specific predefined properties can be of practical 
interest. For example, a portfolio which is least risky among family of 
ones offering the given future return $G_P$. Risk $R(P)$ is usually 
quantified in terms of variance of the time series of historical returns 
\begin{equation}
R (P) = \sigma^2 (P) = {\rm var} \{G_P(j)\}_{j=1}^T.
\end{equation}
According to the classical Markowitz theory, this risk can be related to 
correlations (or covariances) between the time series of individual 
security returns $g_s(j), j=1,...,T$ for the relevant group of securities.

\begin{figure}
\epsfxsize 8cm
\hspace{-1.0cm}
\epsffile{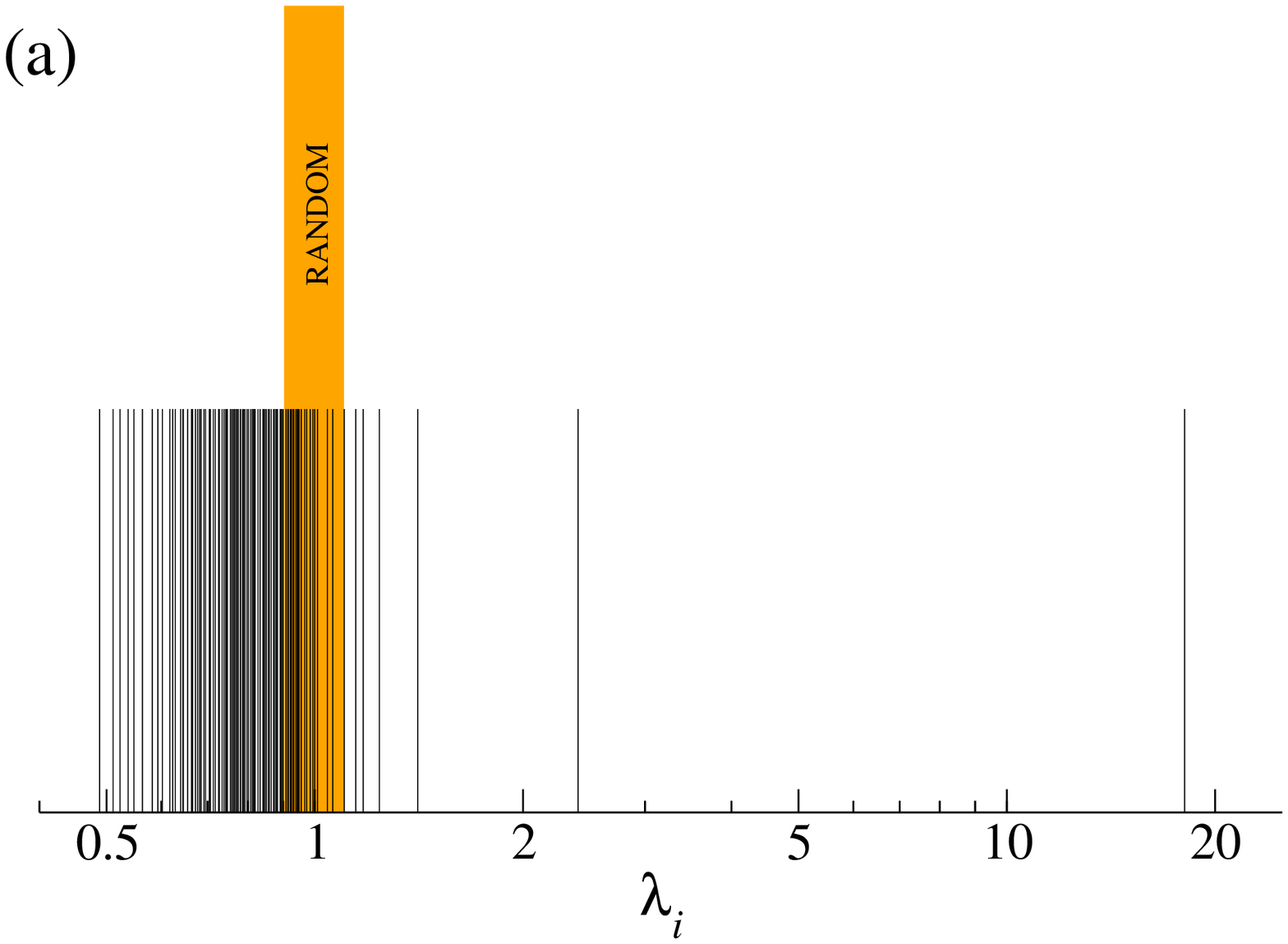}
\hspace{-0.7cm}
\epsfxsize 8cm
\epsffile{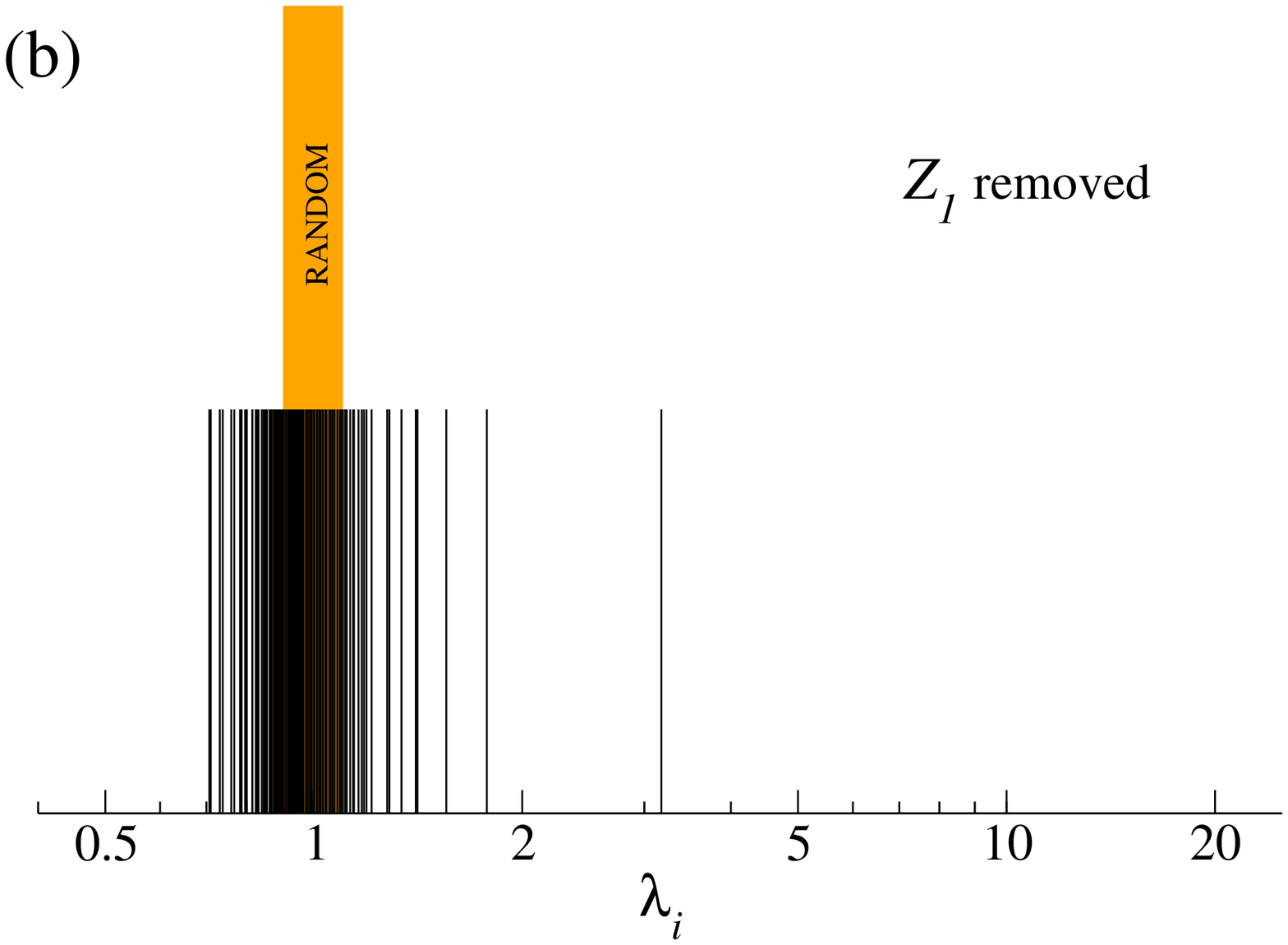}

\vspace{-0.2cm}
\hspace{3.0cm}
\epsfxsize 8cm
\epsffile{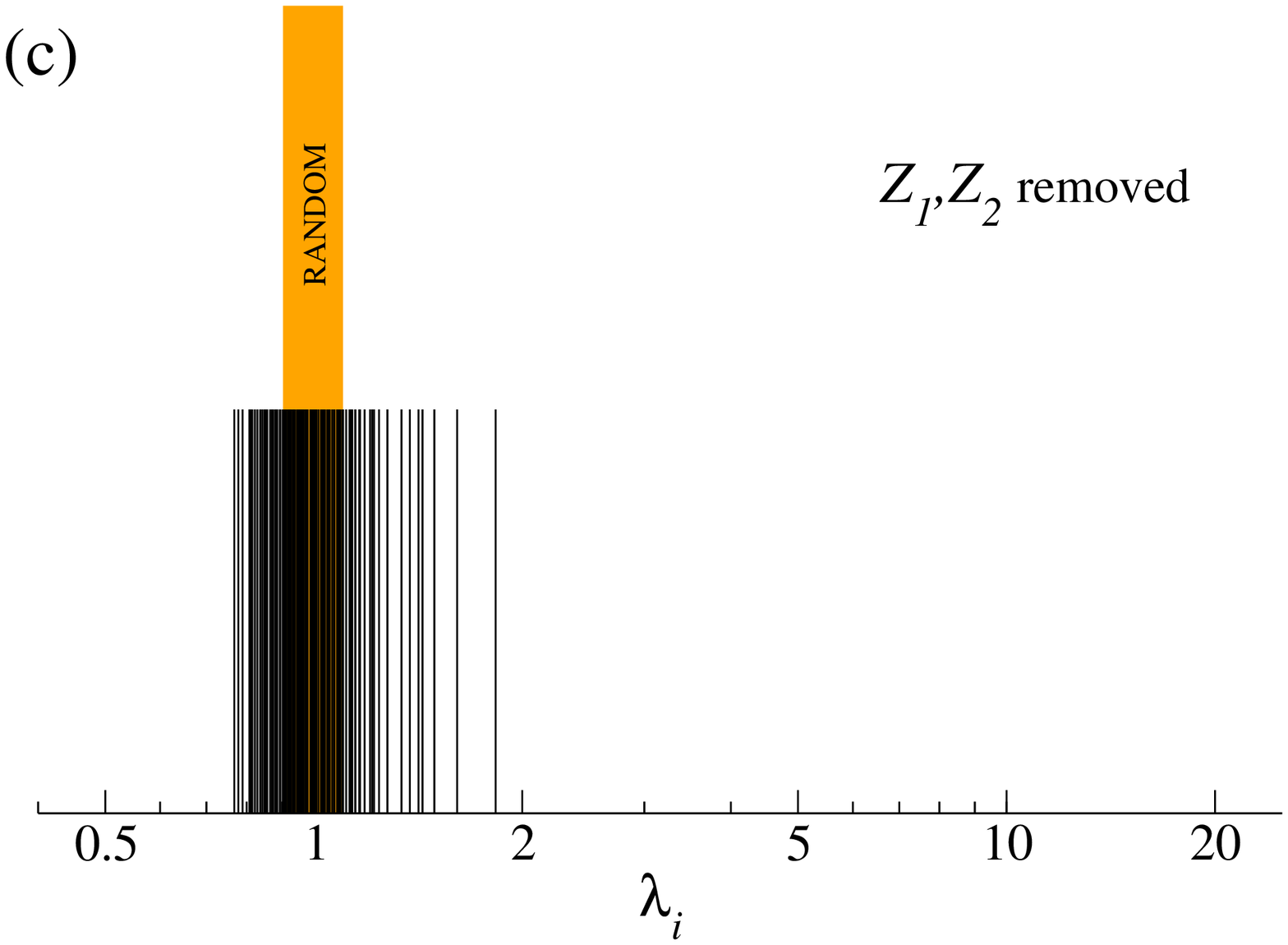}
\caption{Empirical eigenvalue spectrum of the correlation matrix ${\bf C}$ 
(vertical lines), calculated for 100 highly capitalized American 
companies over the period 1998-1999; the eigenvalues of a random Wishart 
matrix with the same $Q$ may lie only within the shaded vertical region 
(a). Eigenvalue spectrum after effective rank reduction of ${\bf C}$, i.e. 
after subtracting the contribution of the most collective eigensignal 
$Z_1$ (b) and the two most collective ones $Z_1$ and $Z_2$ (c).}
\end{figure}

More specifically, let one consider a set of $N$ securities (e.g. stocks)
each represented by a time series of normalized returns $g_s(j),  
s=1,...,N; j=1,...,T$. From these time series an $N \times T$ data 
matrix ${\bf M}$ can be created and then a correlation matrix ${\bf C}$ 
according to the formula 
\begin{equation}
{\bf C} = (1 / T) {\bf M} {\bf M}^{\rm T}. 
\end{equation}
Each element of ${\bf C}$ is obviously the Pearson correlation 
coefficient $C_{m,n}$ between a pair of signals $m$ and $n$. The 
correlation matrix can be diagonalized by solving the eigenvalue problem
\begin{equation}
{\bf C} {\bf x}_i = \lambda_i {\bf x}_i, \ \ i = 1,...,N.
\end{equation}
From the point of view of investment theories, each eigenvector ${\bf 
x}_i$ can be considered as a realization of an $N$-security portfolio 
$P_i$ with the weights equal to the eigenvector components $x_i^{(k)}, 
k=1,...,N$. For a non-degenerate matrix ${\bf C}$, $P_i$ and $P_j$ are 
independent for each pair of their indices, which allows one to choose 
such a portfolio, whose risk is independent of others. 

\begin{figure}
\epsfxsize 8cm
\hspace{-1.0cm}
\epsffile{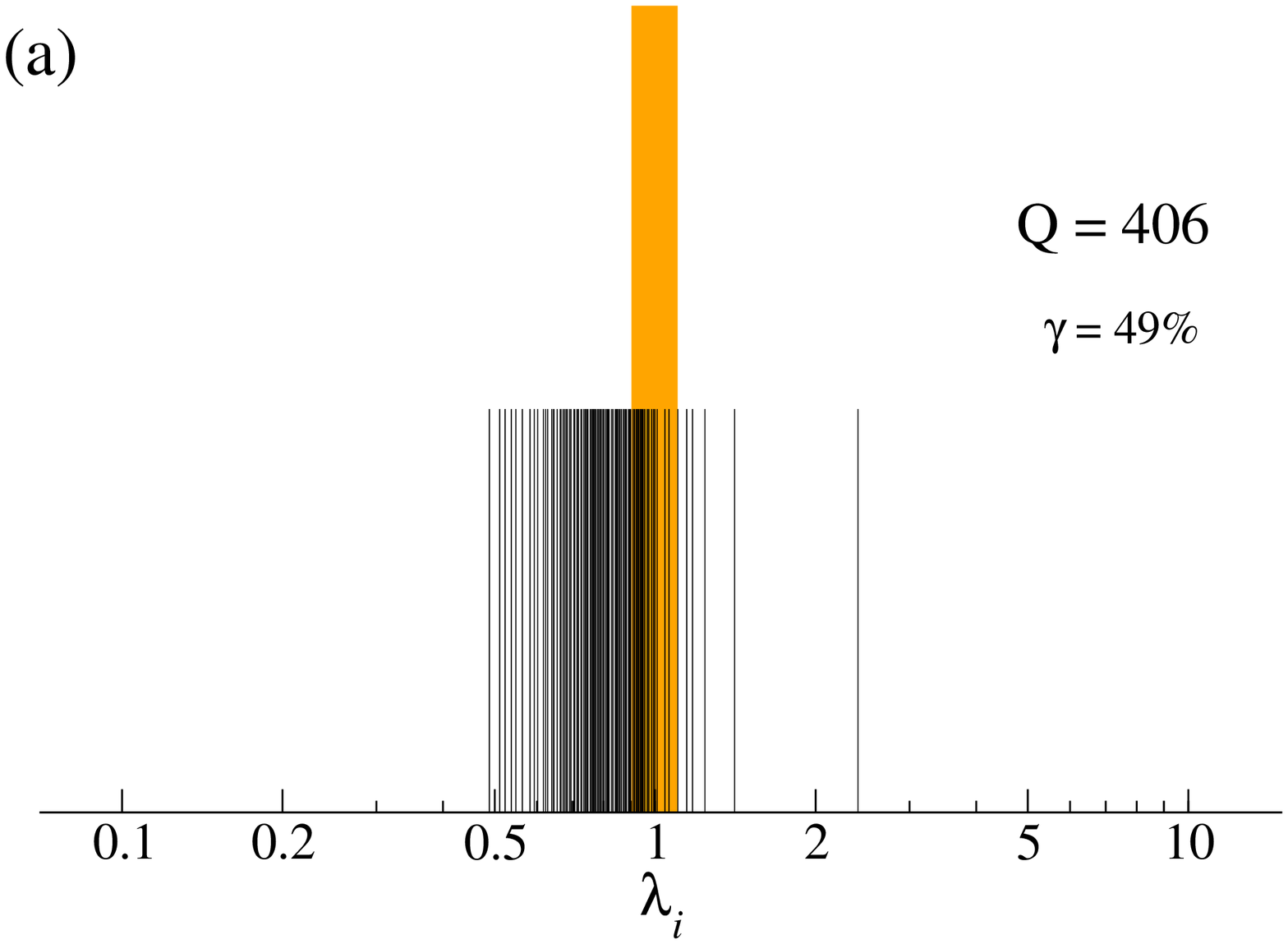}
\hspace{-0.7cm}
\epsfxsize 8cm
\epsffile{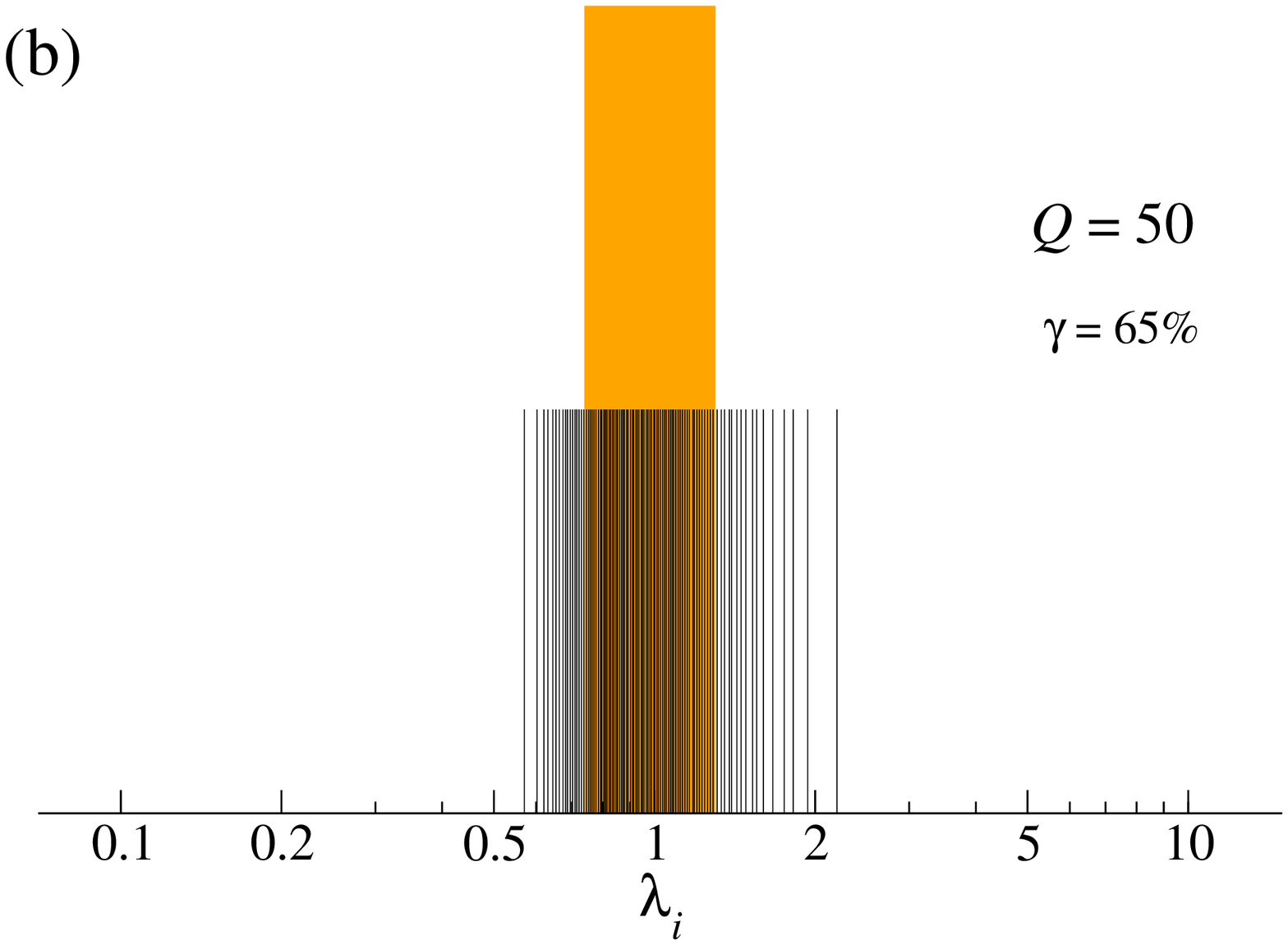}

\vspace{-0.2cm}
\hspace{-1.0cm}  
\epsfxsize 8cm
\epsffile{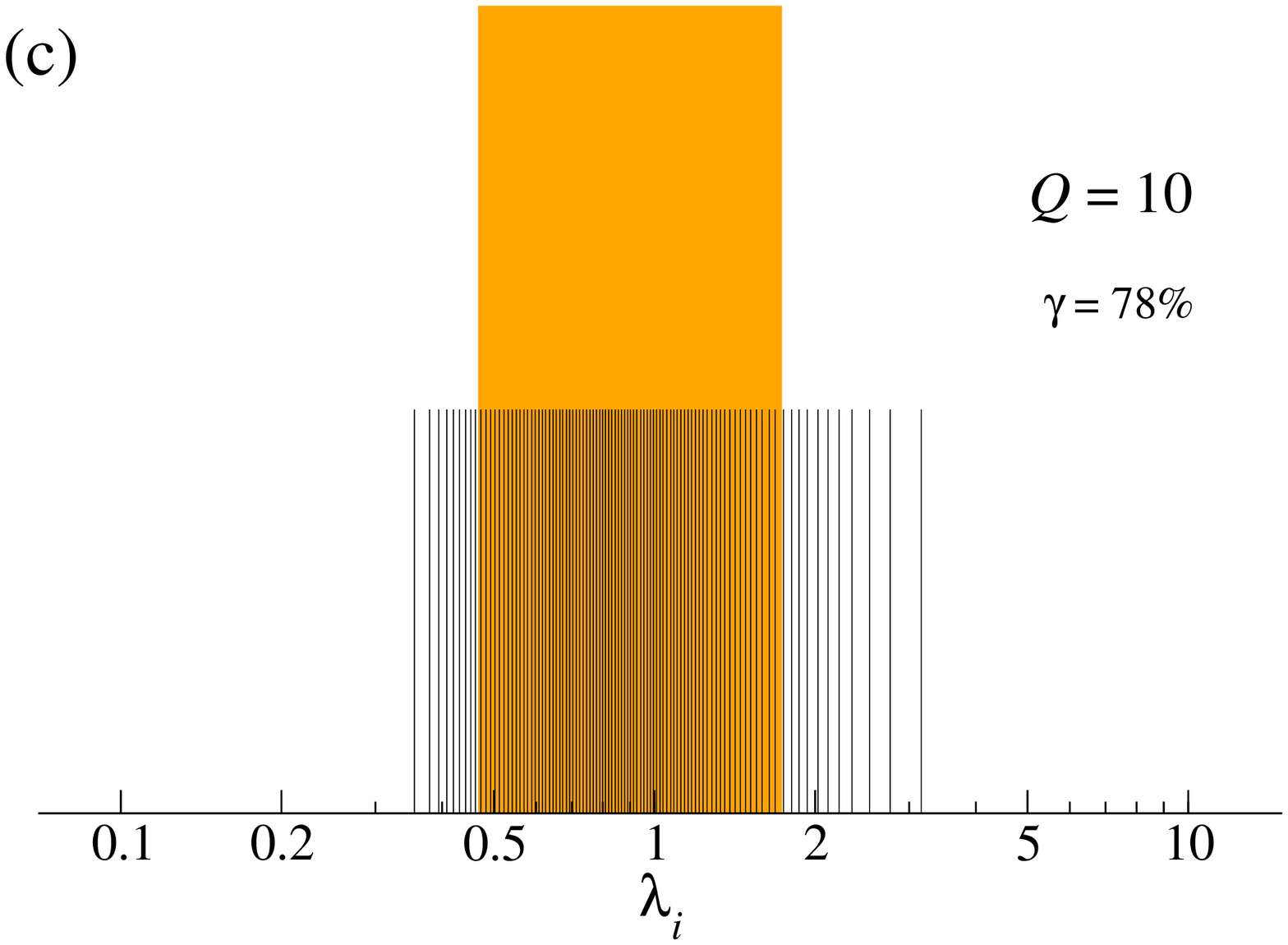}
\hspace{-0.7cm}
\epsfxsize 8cm  
\epsffile{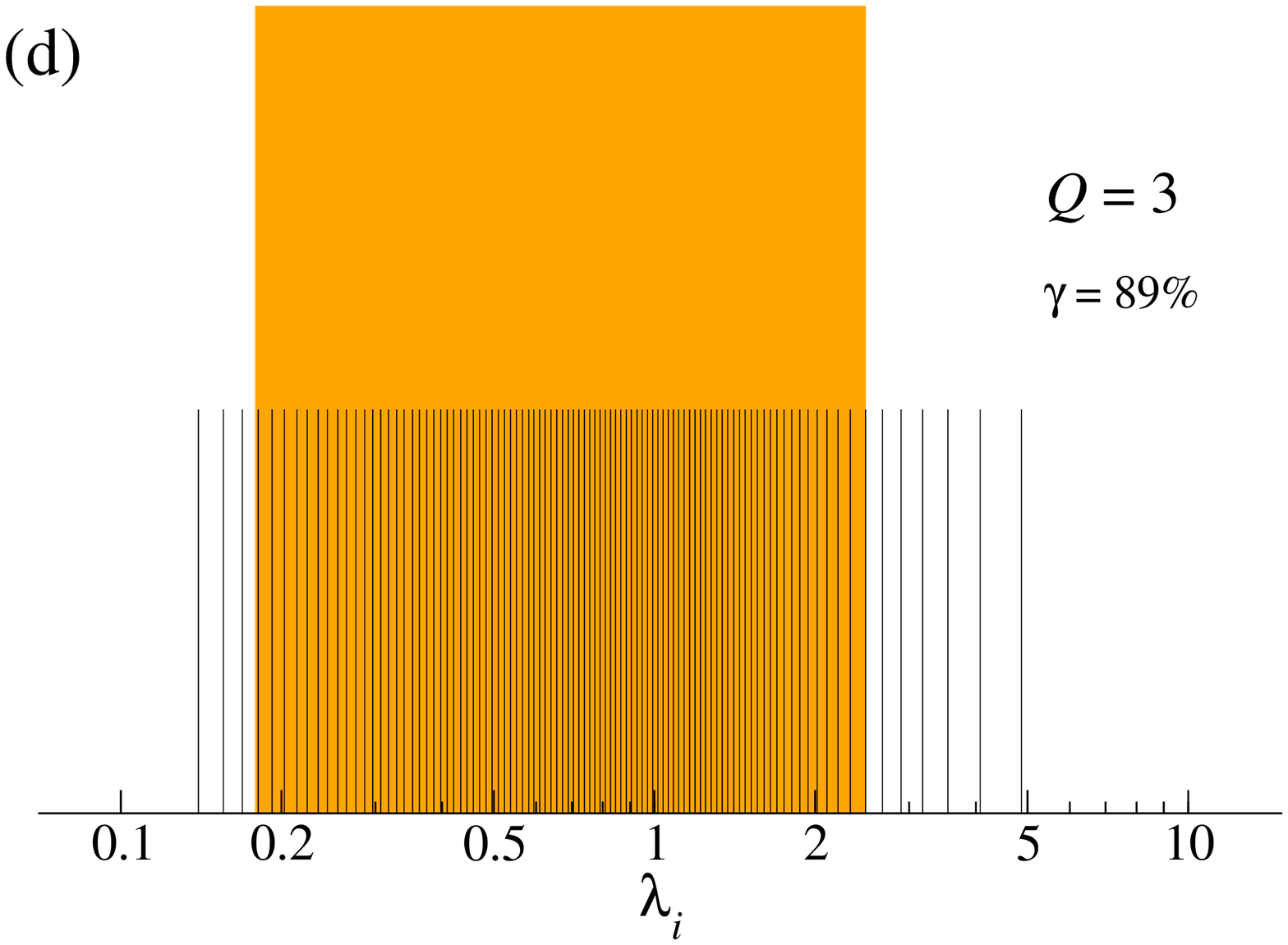}
\caption{Empirical eigenvalue spectrum of correlation matrix ${\bf C'}$ 
(see text) after subtracting the contribution of the two most collective 
components associated with $\lambda_1$ and $\lambda_2$ for four different 
values of $Q$; (a) is the same as Fig.~1(c). Shaded regions correspond to 
RMT predictions for given $Q$.}
\end{figure}

From a dynamical perspective, each eigenvector (and thus portfolio) can be 
associated with the corresponding time series of the portfolio's returns 
by the expression analogous to Eq.(\ref{portreturn})
\begin{equation}
z_i(j) = \sum_{k=1}^N x_i^{(k)} g_k(j), \ \ i = 1,...,N; \ 
j=1,...,T.
\label{eigensignals}
\end{equation}
These time series we shall call the eigensignals $Z_i$ (see
also~\cite{kwapien02,kwapien00} for an alternative realization). One of
the most important properties of such eigensignals is that their risk can
be easily related with the corresponding eigenvalues:
\begin{equation}
R (P_i) = \sigma^2 (Z_i) = {\bf x}_i^{\rm T} {\bf C} {\bf x}_i = \lambda_i,
\label{eigenrisk}
\end{equation}
Thus, the eigenvalue size is a risk measure and, in consequence, the 
larger $\lambda_i$, the larger variance of $Z_i$ and also the larger risk 
of the corresponding portfolio $P_i$.

\section{Results}

\begin{figure}
\epsfxsize 11cm
\hspace{1.5cm}
\epsffile{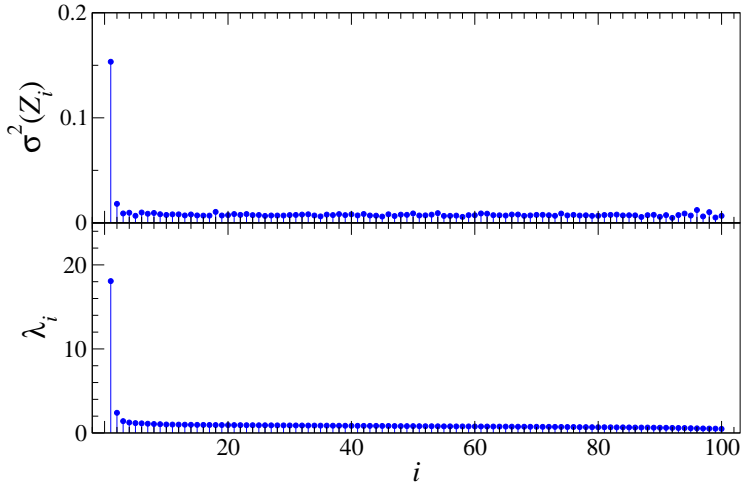}
\caption{Standard deviation $\sigma (Z_i)$ for all eigensignals $Z_i$
(top panel) together with the corresponding eigenvalues $\lambda_i$
(bottom panel). Almost perfect correspondence between both panels is
visible; the difference in units in the vertical axes is due to
eigenvector normalization performed by diagonalization procedure.}
\end{figure}  

We analyzed high-frequency data from the American stock market~\cite{data}
in the period 1 Dec 1997 $-$ 31 Dec 1999. In order to calculate
inter-stock correlations we chose a set of stocks of $N=100$ highly
capitalized companies listed in NYSE or NASDAQ (capitalization $>
\$10^{10}$ in each case). These stocks are frequently traded (0.01-1
transactions/s) and hence the time scale of $\Delta t=5$ min allowed us to
perform a statistically significant analysis; such a short time scale was
also desired because of length of time series (approx. 40,000 data
points). Typically, the portfolio analysis is performed on daily data;
usually this time scale is recommended because of the fact that at this
scale the correlations present at the market are well-developed. However,
restricting the analysis to such long $\Delta t$ seems not to be necessary
in contemporary markets, in which both pieces of information arrive more
frequently and the investors react to them quicker than in the past. This
obviously leads to acceleration of the market time paced by consecutive
transactions. Thus, as recent papers document~\cite{drozdz03,kwapien04},
for highly capitalized stocks which are also among the most frequently
traded ones, the correlations reach their saturation level at the time
horizons much shorter than a day. In case of our data this time horizon
corresponds to about 30 min, while for $\Delta t=5$ min the correlations
associated with $\lambda_1$ are clearly very pronounced and informative
and $\lambda_1$ assumes approx. 2/3 of its saturation level~\cite{kwapien04}.

\begin{figure}
\epsfxsize 13cm
\hspace{0.5cm}
\epsffile{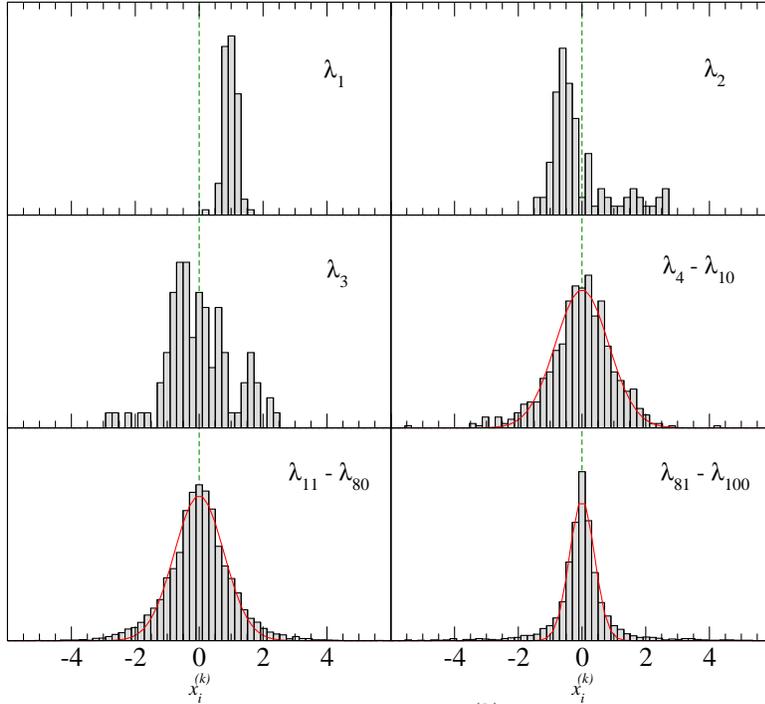}
\caption{Histograms of eigenvector components $x_i^{(k)}$ for different  
eigenvalues $\lambda_i$. A Gaussian is fitted to the empirical histograms
corresponding to the most random eigenvectors.}
\end{figure}

\subsection{Eigenvalue distribution}

The eigenvalue distribution offers a representative and the most
frequently used measure to quantify the characteristics of matrices,
especially in the context of relating them to RMT. Let us therefore start
presentation of the results with the eigenvalue spectrum of the
correlation matrix. Figure 1(a) shows all 100 eigenvalues distributed
along the horizontal axis, denoted by vertical lines. The largest
eigenvalue $\lambda_1 \simeq 18$, repelled from the rest of the spectrum,
describes the collective eigenstate which can be identified with the
market. $\lambda_2 \simeq 2.4$ absorbs some collectivity either, but its
magnitude is by an order of magnitude smaller than $\lambda_1$ and can be
related with some branch-specific factor (the same applies to a few next
eigenvalues). Figure 1(a) displays that only a small fraction of
$\lambda_i$'s falls within the RMT region defined by its bounds:  
$\lambda_{\rm min}^{\rm max} = 1 + 1 / Q \pm 2 / \sqrt Q$ (shaded vertical
region in Figure), where $Q = T/N \simeq 406$~\cite{sengupta99}.  
However, due to the fact that ${\rm Tr}{\bf C}=100$, the existence of
strong collective components can effectively supress the noisy part of the
${\bf C}$ eigenspectrum, shifting smaller eigenvalues towards zero.
Therefore, in order to correct for this effects it is recommended to
remove the market factor $Z_1$ from the data~\cite{plerou02}. This can be 
done by means of the least square fitting of this factor represented by 
$z_1(j)$ to each of the original stock signals $g_k(j)$:
\begin{equation} g_k(j)=\alpha_k +
\beta_k z_1(j) + \epsilon_k^{(1)}(j),
\end{equation} 
where $\alpha_i,\beta_i$ are parameters, and then we can construct a new
correlation matrix ${\bf C}^{(1)}$ from the residuals
$\epsilon_k^{(1)}(j)$ (e.g. ref.~\cite{laloux99,plerou02}). Now
significantly more eigenvalues fall within the shaded RMT region as Figure
1(b) documents. This can be done once again and the $\lambda_2$ component
can also be removed leading to the eigenspectrum presented in Figure 1(c).
In fact, now many more eigenvalues ($\gamma=49$\%) overlap with the RMT
interval $<\lambda_{\rm min},\lambda_{\rm max}>$, though, interestingly,
this value is qualitatively different from results presented earlier
in~\cite{laloux99,plerou02,kwapien02} where vast majority of the
eigenvalues was inside the RMT bounds.

In order to shed some light on this problem we note that in both cited
works the parameter $Q$ was much smaller than in our case: $Q=4.1$ and 6.4
in ref.~\cite{plerou02} and $Q=3.2$ in~\cite{laloux99} and therefore the
RMT spectrum was respectively wider. By manipulating the $Q$ value for our
data (we divide the time series into windows of length $T'$ with
predefined $Q'=T'/N$, then we calculate ${\bf C}'$ for each window and
average its eigenspectrum over all the windows) we obtain the correlation
matrix eigenspectrum which can be compared with the original one for the
undivided time series ($Q=406$). Since in each case the average
$\lambda_1$ is strongly repelled (and its value is approximately the same
as in Fig.~1(a)), we follow the earlier procedure and remove the
collective components related to both $\lambda_1$ and $\lambda_2$ for each
window before averaging the resulting eigenspectra. Figure 2 shows such
decollectified average eigenspectra for four different values of $Q$. What
is immediately evident, the wider the shaded RMT region, the more
eigenvalues it overlaps with. For the smallest presented $Q=3$ as much as
$\gamma=89$\% eigenvalues fall within the RMT realm which is compatible
with $\gamma=94$\% from ref.~\cite{laloux99}. Naturally, as we verified by
an explicit calculation, this observation obtained for $\Delta t=5$ min
qualitatively remains unaltered if we pass to longer time scales e.g.
$\Delta t=60$ min, despite the fact that the corresponding time series
shorten and the maximum available $Q$ decreases. We conclude that for a
typical realization of $Q<10$ in practical applications (usually large $N$
and relatively small $T$ as it happens for daily data), only the largest
eigenvalues are able to deviate from the RMT predictions, while the other
eigenvalues possibly carrying some more subtle correlations may be forced
to spuriously merge with the random bulk. This purely statistical effect
suggests that the random part of the ${\bf C}$ eigenspectrum can in fact
comprise non-random components which can be discerned from noise only if
one uses data with larger $Q$. This is in favour of using data also with
frequencies higher than the daily one as a tool for denoising ${\bf C}$.
(It is noteworthy that a parrallel effect of shifting the small non-random
eigenvalues ($\lambda_i \approx \lambda_{\rm min}$) into the RMT interval
has been presented recently in ref.~\cite{utsugi04}.)

\begin{figure}
\epsfxsize 13cm
\hspace{0.5cm}
\epsffile{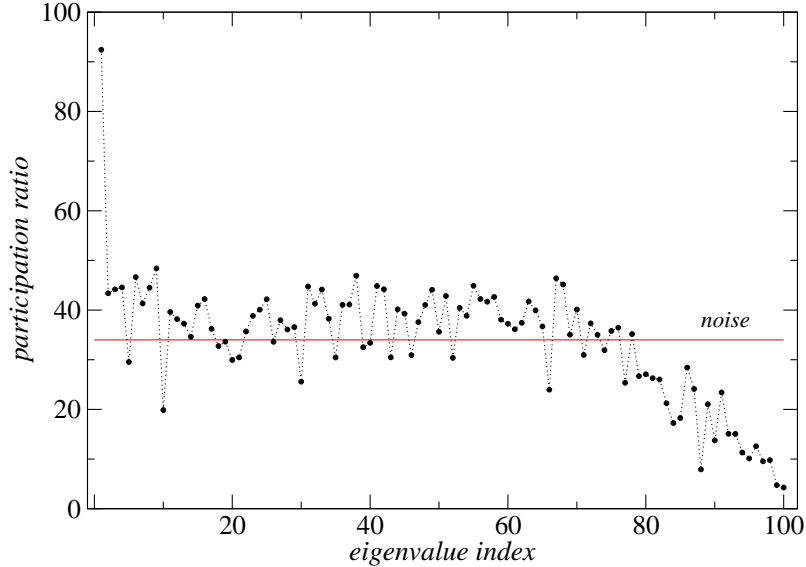}
\caption{Reciprocal of inverse participation ratio (Eq.(\ref{ipr})) for
all 100 eigenvectors of the correlation matrix. Noise level for random
Gaussian case is also presented.}
\end{figure}

\subsection{Eigenvector properties}

Figure 3 illustrating the strong relation between the risk and the
eigenvalues (Eq.(\ref{eigenrisk})), displays the eigensignal variance
calculated for each $Z_i$ (upper panel) and all the eigenvalues of ${\bf
C}$ (lower panel).  Essentially no significant qualitative difference
between these two quantities can be found, exactly as expected. Apart from
the eigenvalue spectrum, RMT offers useful predictions regarding
distribution of the eigenvector components for a completely random matrix,
which assumes the form of the Porter-Thomas (Gaussian)  distribution. In
contrast, in the case of a collective non-random eigenvector, there exists
some kind of vector localization or delocalization. Figure 4 presents
distributions of the components for typical eigenvectors of our matrix and
for a few specific cases. The eigenvector corresponding to the largest
eigenvalue $\lambda_1$ is completely delocalized because all its
components are roughly the same and the associated distribution is
centered at 0.1. This is standard situation and in evolution of the stock
prices this eigenvector represents the market factor. Also non-random is
the eigenvector for $\lambda_2$ with the components distribution still far
from Gaussian. A trace of randomness however occurs already for
$\lambda_3$ and is clear for a bulk of eigenvectors in the next panel of
Figure 4. On the other hand, for the extremely small eigenvalues,
localization can be perfectly seen.

\begin{figure}
\epsfxsize 13cm
\hspace{0.5cm}
\epsffile{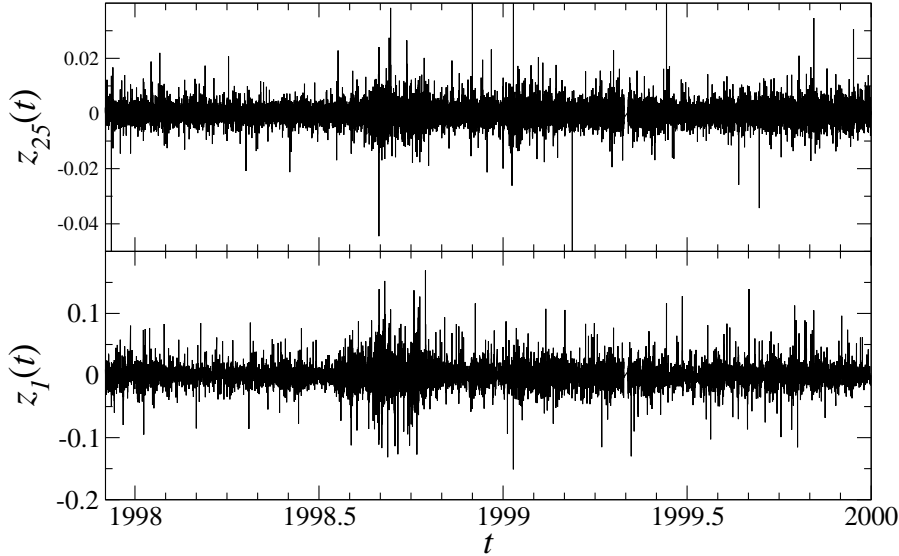}
\caption{Time series of the eigensignals for $\lambda_1$ (top) and
$\lambda_{25}$ (bottom). Note different scales in vertical axes of both 
panels.}  
\end{figure}

One of the key properties of the eigenvector (and thus also the
eigenstate $Z_i$) is the effective number of its large components. A 
related measure is inverse participation ratio
\begin{equation}
I_i=\sum_{k=1}^N (x_i^{(k)})^4, \hspace{0.5cm} k=1,...,N
\label{ipr}
\end{equation}
and its reciprocal $1/I_i$ (``participation ratio''). Figure 5 presents
this latter quantity calculated for all the eigenvectors together with its 
value $N/3$ for random case in which $x_i^{(k)}$ are taken independently 
from normal distribution. For $i=1$ almost all the companies contribute to 
the corresponding eigenvector $Z_1$, which justifies treating this 
eigenvector as the market factor. The eigenvectors for a few smaller 
eigenvalues show also slightly higher number of participating companies 
than for the random case but, in contrast, the eigensignals associated 
with the smallest 20 eigenvalues allow one to characterize them as the 
components related to only few stocks. The rest of the eigenvectors seem 
to be random, with small deviation from the predicted value of $N/3$ 
probably due to the existence of fat tails of the returns distributions.

\begin{figure}
\epsfxsize 13cm
\hspace{0.5cm} 
\epsffile{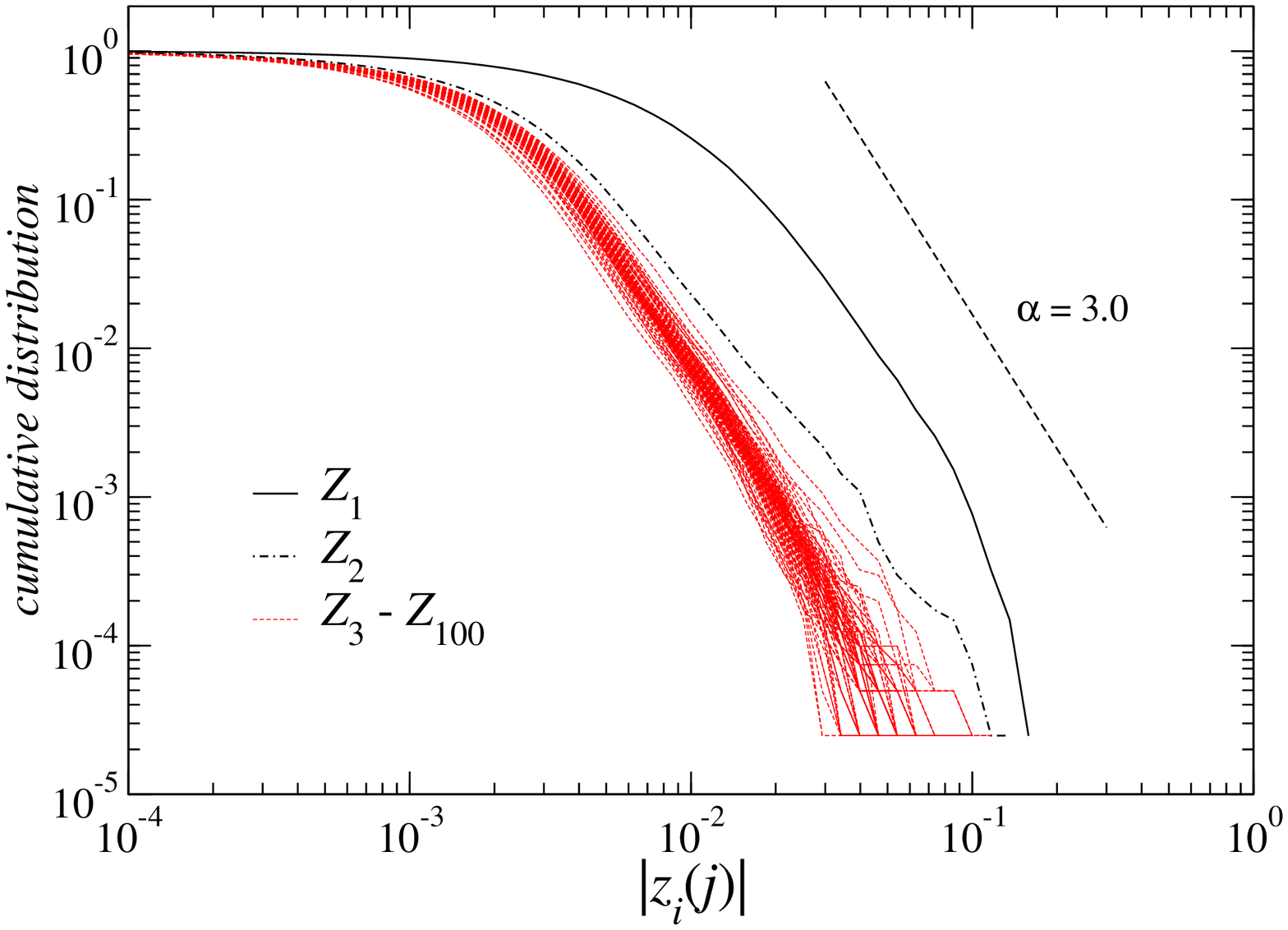}

\hspace{0.5cm}
\epsfxsize 13cm
\epsffile{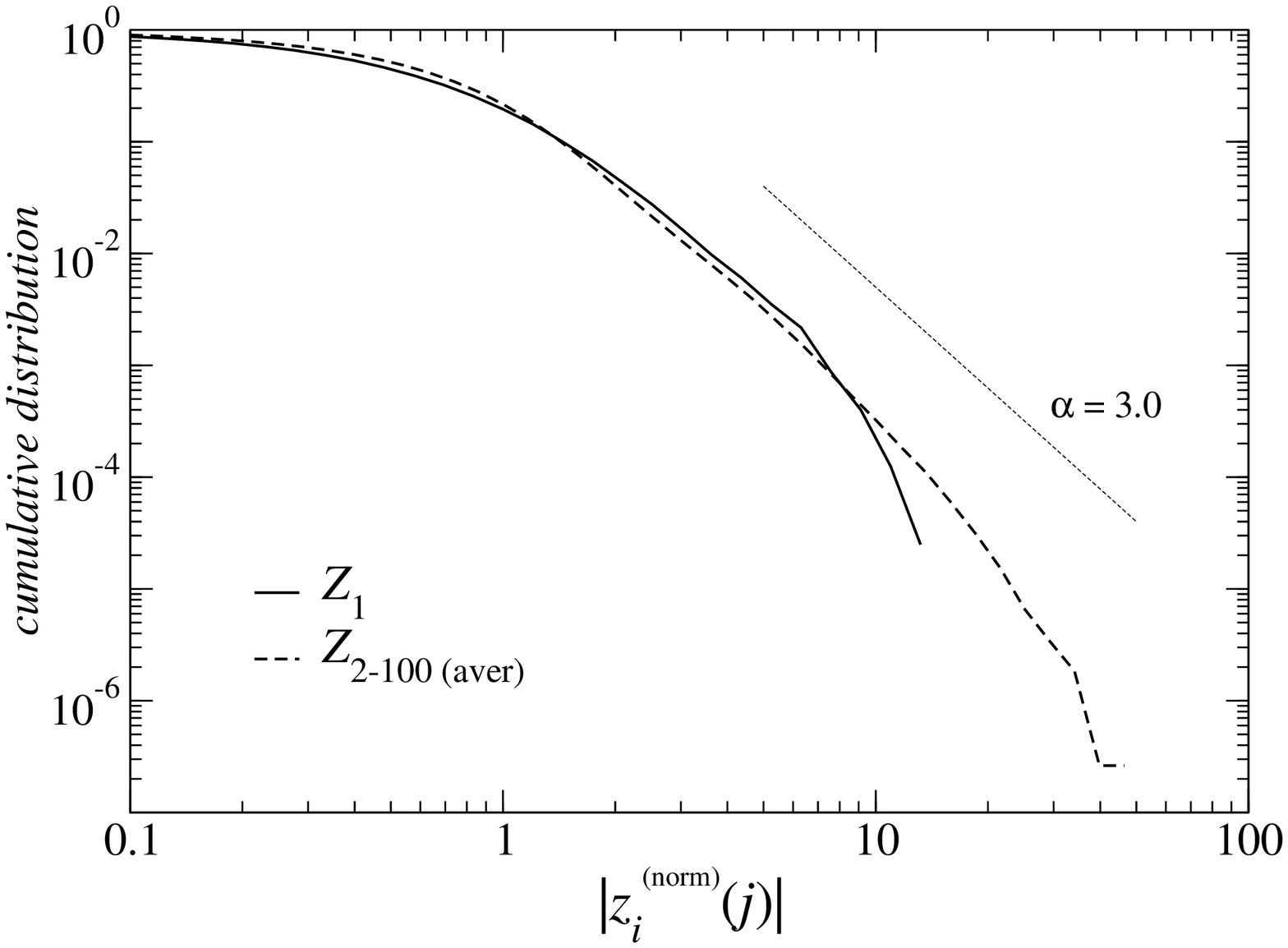}
\caption{(a) Cumulative distribution functions for the eigensignal
returns for all $Z_i$ ($i = 1,...,N$). Distributions for $Z_1$ (thick
solid) and $Z_2$ (dot-dashed) are distinguished. Inverse cubic power law
($\alpha=3.0$) is also denoted by dashed line. (b) The c.d.f. for the
eigensignal returns after normalization to unit variance and averaging
over $Z_i, i=2,...,100$ (dashed) together with c.d.f. for $Z_1$
(solid).}
\end{figure}

Figure 6 presents the time series of the eigensignal returns
$z_1(j)$ calculated according to Eq.(\ref{eigensignals}) for
$\lambda_1$ and $\lambda_{25}$. Curiously, if one compares both series
visually, forgetting the difference in vertical axis, it could be quite
difficult to point out which of the two corresponds to the most collective
eigenstate. Both eigensignals are nonstationary with likely extreme
fluctuations and both of them reveal also volatility clustering. Thus, one
can infer that there are statistical properties which are invariant under
change of the eigenstates with only minor differences between collective 
and noisy eigenstates.

In agreement with Figure 3, c.d.f. of the eigensignal returns (Figure
7(a)) show that $Z_1$ is characterized by much broader distribution than
other eigensignals, and that the same, but to a lesser extent, is true
also for $Z_2$. It is interesting that, except for $Z_1$, all the
eigensignals are associated with the distributions with the power law
scaling in tails, almost exactly like for the original stock returns (see
e.g.~\cite{drozdz03,plerou99b,kwapien03,gabaix03}). Only $Z_1$ presents
different behaviour: a short range of power law scaling and significant
deviation from this behaviour for $|z_1| > 5\sigma$. This can be even more
convincing if all the signals are normalized to unit variance (Figure
7(b)); the power law slope with $\alpha\simeq 3$ is typical for all the
eigensignals (although for a considerably larger $\Delta t$ exactly this
kind of scaling may not be observed~\cite{drozdz03}, the eigensignal
c.d.f.s also preserve tails of the individual stock returns c.d.f.s for
the same $\Delta t$).  The non-typical shape of c.d.f. for $Z_1$ in Figure
7(b) can originate from the fact that this eigensignal is composed as an
average of almost 100 individual stocks, while on average 1/3 of this
number of stocks contribute to other eigensignals; this is why Central
Limit Theorem leaves its fingerprints presumably on $Z_1$. However, this
cannot be considered as a rule, because for significantly larger time
scales (e.g. $\Delta t=60$ min) or for some other groups of stocks for
which the cross-correlations are more intensive such a peculiarity of the
$Z_1$'s c.d.f. might not be observed.  Noteworthy to mention is that the
inter-stock correlations responsible for a transfer of
scaling~\cite{kwapien03} from the stock returns to the eigensignal
distributions in Figure 7 are mainly linear (thus detectable by the
correlation matrix). However, the very fact that scaling exists at $\Delta
t=5$ min or at longer time scales is an effect of highly nonlinear
temporal correlations present in the stock returns (the volatility
autocorrelations etc.) which may also be detectable in the eigensignals
(see the next subsection).

\begin{figure}
\epsfxsize 13cm
\hspace{0.5cm}
\epsffile{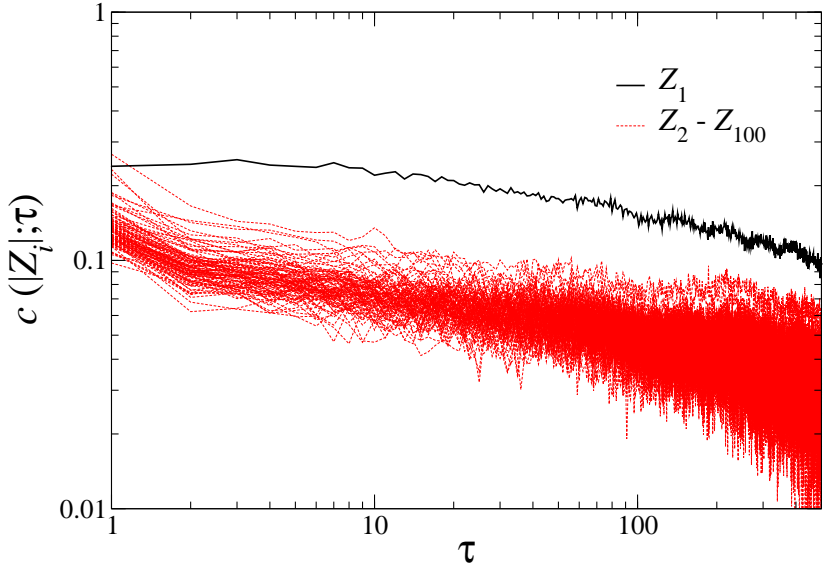}
\caption{Volatility autocorrelation function $C(|Z_i|,\tau)$ for
$i=1,...,100$. For each eigensignal $Z_i$, daily trend was removed by the
standard detrending procedure (see~\cite{gopi99,kwapien05}).}
\end{figure}

\subsection{Temporal correlations}

Now let us look at temporal correlation properties of the eigensignals.
Figure 8 shows the volatility autocorrelation function $c(|Z_i|;\tau)  
\equiv c(|z_i(j)|,|z_i(j+\tau)|)$ for all the eigensignals $Z_i$. The
difference between $Z_1$ and the rest of signals is pronounced and
resembles the corresponding difference in the case of the eigensignals
variance (Figure 3). In fact, memory in $Z_1$ is about two orders of
magnitude longer than for the other eigensignals (for $m>1$:  
$c(|Z_i|;\tau) \simeq c(|Z_1|;10^2*\tau)$). However, it cannot be said
that $Z_1$ (the market component) absorbs all the memory of the evolution
of stock prices. The volatility autocorrelation function for the other 
eigensignals decay very slowly in time as well.

Another interesting sort of nonlinear correlations are cross-correlations
between returns and volatility $c(Z_i,|Z_i|;\tau)$. It has been observed
that past returns imply negatively correlated future moves in volatility,
the so-called leverage effect~\cite{bouchaud01,masoliver02,eisler04}.  
Figure 9 shows the cross-correlation function $c(Z_i,|Z_i|;\tau)$ for the
returns and the volatility for all the eigensignals. The leverage effect
can be detected easily only for $i=1,2,3$; all other eigensignals do not
reveal it.  For $i=1$ and $i=2$ the correlations are qualitatively the
same despite the earlier-presented differences between the eigensignals.
For $i=3$ negative correlations do not exist for the smallest values of
$\tau$, while they are the strongest for $\tau>100$. In order to carry out
this calculation we took signals for $\Delta t=20$ min, because analogous
signals at smaller time scales are too noisy to show the strong leverage
effect. Interestingly, for the original signals $g_s$ corresponding to
individual stocks $s$, the leverage effect is much weaker and even
difficult to observe at all.

\begin{figure}
\epsfxsize 13cm
\hspace{0.5cm}
\epsffile{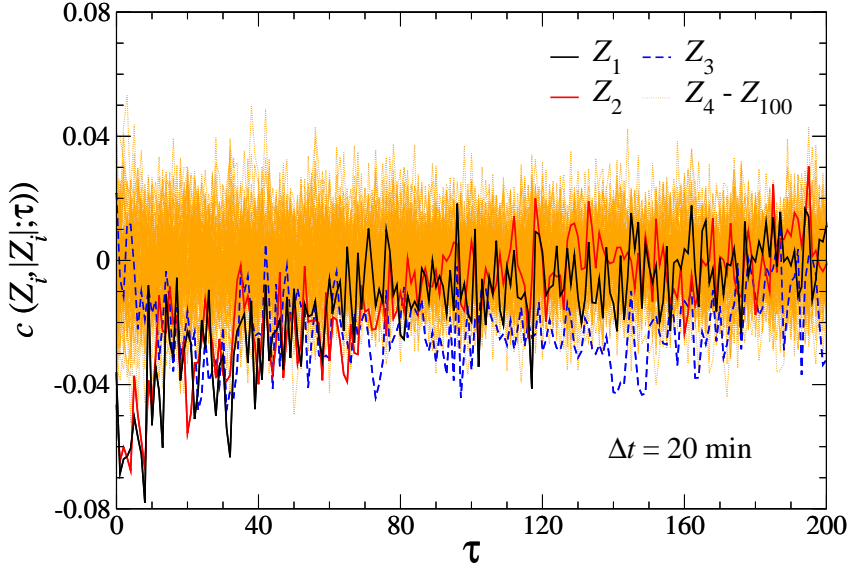}
\caption{Cross-correlation function $c(Z_i,|Z_i|;\tau)$ for time series
of returns $Z_i$ and of volatility $|Z_i|$ calculated for all eigensignals
separately. The resulting functions for the eigensignals corresponding to
three largest eigenvalues are distinguished by thick lines.}
\end{figure}

\subsection{Multifractal characteristics}

One of the consequences of the broad probability distributions (Figure 7)  
and of the nonlinear correlations (Figure 8 and 9) is multifractal
character of signals, i.e. the existence of a continuous spectrum of
scaling indices $f(\alpha)$. It has already been shown in numerous works
that stock returns form signals which are multifractal both on daily and
on high-frequency time scales~\cite{pasquini99,ivanova99,bershadskii03,%
matteo04,fisher97,vandewalle98,bershadskii99,matia03,oswiecimka05a,kwapien05}.  
This behaviour can be even modeled with a good agreement if one introduces
statistical processes based on multiplicative
cascades~\cite{eisler04,mandelbrot97,calvet97,lux03a,lux03b}. Due to the
fact that, by definition, the eigensignals are calculated as a sum of
stock returns, the multifractality of their components can be transferred
to the resulting eigensignal. Therefore one can expect that at least some
of $Z_i$'s are also multifractal;  there can exist differences in their
singularity spectra because of different stock compositions for different
$i$'s, though. Owing to superiority of Multifractal Detrended Fluctuation
Analysis (MFDFA)~\cite{peng94} over Wavelet Transform Modulus Maxima
(WTMM) method~\cite{arneodo95} in the case of financial
data~\cite{oswiecimka05b}, we applied former one to our time series in
order to calculate the singularity spectra $f(\alpha)$.

\begin{figure}
\epsfxsize 13cm
\hspace{0.5cm}
\epsffile{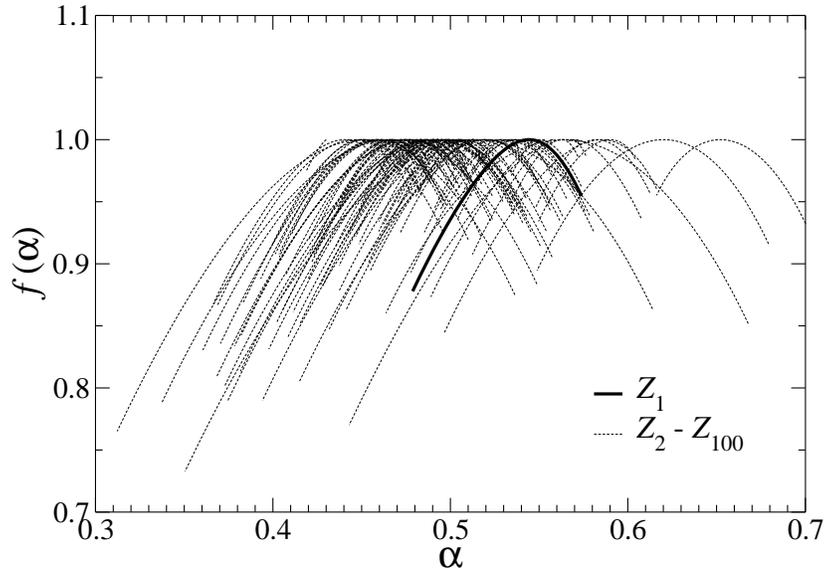}

\hspace{0.5cm}
\epsfxsize 13cm
\epsffile{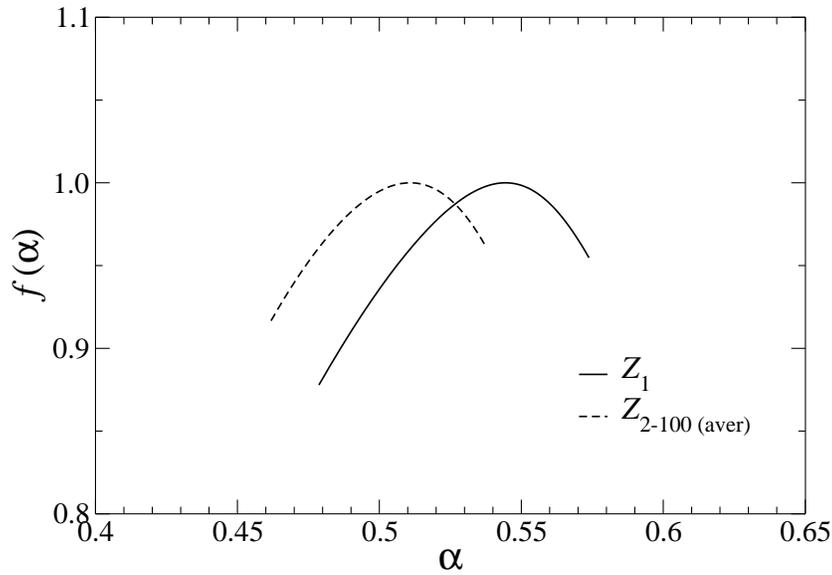}
\caption{(a) Singularity spectra $f(\alpha)$ for eigensignals
corresponding to all correlation matrix eigenvalues $\lambda_i$. Spectrum
for the eigensignal $Z_1$ associated with the largest eigenvalue is
denoted by a thick solid line; (b) singularity spectrum for $Z_1$ (solid)
compared with the average spectrum for all other eigensignals $Z_i, \ i =
2,...,100$ (dashed).}
\end{figure}

We start from our eigensignal $i$ represented by the time series 
$z_i(j)$ of length $N_{es}$ and estimate the signal profile
\begin{equation}
Y(j) = \sum_{k=1}^j{(z_i(k)-<z_i>)}, \ j = 
1,...,N_{es}
\end{equation}
where $<...>$ denotes averaging over $z_i(k)$. In the next step 
$Y$ is divided into $M_{es}$ segments of length $n$ ($n < N_{es}$) 
starting from both the beginning and the end of the time series so that 
eventually there are $2 M_{es}$ segments. In each segment $\nu$ we fit a 
$l$-th order polynomial $P_{\nu}^{(l)}$ to the data, thus removing a 
local trend. Then, after calculating the variance
\begin{equation}
F^2(\nu,n) = \frac{1}{n} \sum_{k=1}^n \{Y[(\nu-1) n+k] -
P_{\nu}^{(l)}(k)\}^2
\end{equation}
and averaging it over $\nu$'s, we get the $q$th order fluctuation
function
\begin{equation}
F_q(n) = \bigg\{ \frac{1}{2 M_{es}} \sum_{\nu=1}^{2 M_{es}} 
[F^2(\nu,n)]^{q/2}\bigg\}^{1/q}, \ \ q \in \mathbf{R}
\end{equation}
for all values of $n$. The most important property of $F_q(n)$ is that for 
a signal of the fractal character it obeys a power-law functional 
dependence on $n$:
\begin{equation}
F_q(n) \sim n^{h(q)},
\label{scaling}
\end{equation}
at least for some range of $n$. As a result of complete MF-DFA procedure 
we obtain a family of generalized Hurst exponents $h(q)$, which form a 
decreasing function of $q$ for a multifractal signal or are independent of 
$q$ for a monofractal one. A more convenient way to present the 
fractal character of data graphically is to calculate the singularity 
spectrum $f(\alpha)$ by using the following relations:
\begin{equation}
\alpha=h(q)+q h'(q) \hspace{1.0cm} f(\alpha)=q [\alpha-h(q)] + 1.
\label{singularity}
\end{equation}

\begin{figure}
\epsfxsize 13cm
\hspace{0.5cm}
\epsffile{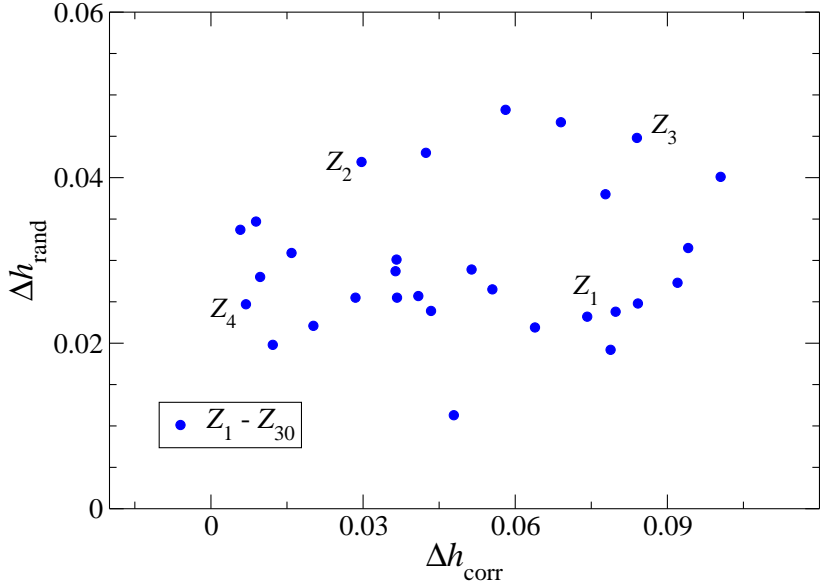}
\caption{$\Delta h_{\rm corr}$ vs. $\Delta h_{\rm rand}$ for
eigensignals corresponding to 30 largest eigenvalues. Symbols denoting
$Z_1,...,Z_4$ are labeled.}
\end{figure}   

We computed the $f(\alpha)$ spectra with MF-DFA for all time series $Z_i$
and for $\Delta t=5$ min (shorter time scales are too noisy, while longer
ones are represented by too short series for a reliable estimation of
$f(\alpha)$) and the corresponding results are shown in Figure 10(a). Wide
spectra prove that all $Z_i$'s are multifractal with only minor
differences in widths of the spectra for different eigensignals. Although
positions of maxima of the spectra vary, there is no significant
$i$-dependence both in the positions and in shape of the spectra. It is
interesting that even the most collective and correlated eigensignal for
$\lambda_1$ (the solid, distinguished line in Figure 10(a)) does not
develop spectrum which could deviate from the typical one. This is even
more evident if we compare the spectrum for $Z_1$ with the average
spectrum calculated from all other $Z_i$'s in Figure 10(b). Such 
similarity
of multifractal properties of the eigensignals representing completely
different correlation structure and different p.d.f.'s can suggest that
the eigensignals corresponding to a few largest $\lambda_i$'s and the ones
corresponding to the bulk of the eigenspectrum are in fact much more
similar to each other than it is usually assumed. This also bears a 
serious concern regarding the justification of treating the ``noisy'' 
eigenstates as completely random without any information content.

As we already know the autocorrelations of volatility $|Z_i|$ for $i=1$
are different from those for $i>1$ (Figure 8), while the normalized
returns have similar c.d.f.'s for all $i$. Now we would like to compare 
the fractal properties of different eigensignals in order to separate the 
two sources of multifractality: the broad distributions of returns and the
correlations, and to compare the singularity spectra $f(\alpha)$ for these
sources. We follow the idea of~\cite{kantelhardt02} and we study 
variability of generalized Hurst exponents $h(q)$ for the actual and the 
reshuffled eigensignals. If we denote the generalized Hurst exponent for 
the randomized signal by $h_{\rm rand}(q)$, its correlation counterpart 
reads
\begin{equation} 
h_{\rm corr}(q):=h(q)-h_{\rm rand}(q).
\end{equation}
Variability of $h(q)$ can be expressed by the difference 
\begin{equation}
\Delta h:=h(q_{\rm min})-h(q_{\rm max}) 
\end{equation} 
and, analogously, the variability of $h_{\rm rand}(q)$ and $h_{\rm
corr}(q)$. Each symbol in Figure 11 presents positions of the eigensignals
$Z_i$ in ($\Delta h_{\rm corr},\Delta h_{\rm rand}$) coordinates.
Eigensignals corresponding to the four largest eigenvalues are labelled.
The higher value of $\Delta h_{\rm rand}$, the richer the multifractal
behaviour due to the fat-tailed probability distributions of returns.
Analogously, high values of $\Delta h_{\rm corr}$ indicate strong
contribution of the temporal correlations. These temporal correlations
implying the observed multifractal behaviour of signals have to be
strongly nonlinear: our earlier calculations~\cite{kwapien05} showed that 
even the volatility autocorrelations that are nonlinear in returns are not 
capable of producing this phenomenon. Furthermore, it can be easily 
inferred from Figure 11 that there is nothing characteristic in the 
positions of the symbols related to $Z_{1,...,4}$. Results collected in 
Figure 10(b) and Figure 11 indicate that the multifractal analysis cannot 
point out any essential differences between the collective eigenstates and 
the noisy ones.

\section{Conclusions}

We analysed the eigensignals corresponding to different eigenvalues of the
empirical correlation matrix calculated for the 100 highly-capitalized
American companies. From a practical point of view, these eigensignals
represent temporal evolution of specific portfolios defined by the
corresponding eigenvectors of ${\bf C}$. We showed that despite the
important differences in interpretation of different eigensignals, $Z_i$'s
for the collective and for the noisy eigenvalues can reveal different or
similar statistical properties depending on a particular quantity. What
differs most is risk expressed by variance of an eigensignal, which is
very high for the most collective $Z_1$, is also significant for $Z_2$,
and is much smaller for the rest of the $Z_i$'s. This is closely related
to the eigenvector properties, which are different for highest and other
$\lambda_i$'s. The risk is also related to the width of the returns
distributions, preventing the eigensignals for small eigenvalues from high
fluctuations and the associated portfolios from large losses. The second
group of quantities which reveal different values for different $Z_i$'s
are nonlinear correlations, both those in volatility and those between
returns and volatility. On the other hand, there are properties that
remain unaltered when going from small to large values of $i$. The most
interesting one is the multifractality of all eigensignals which is
surprisingly roughly the same for different $Z_i$'s. Curiously, this
happens even if the above-mentioned correlations vary among the
eigensignals. One of possible sources of this can be similar shape of
tails of the probability distributions, which after normalizing the
eigensignal returns show the inverse cubic scaling for all $i$. Both the
fat-tailed distributions of returns and the multifractal character of
eigensignals for each eigenvalue lead to a conclusion that the noisy
eigenstates might not be so random as they are usually regarded by their
relation to RMT. Rich multifractal dynamics of the eigensignals
corresponding to even the random part of the eigenvalue spectrum suggests
that strong nonlinear correlations are present in the temporal evolution
of each portfolio giving it significance which exceeds pure noise. This
conclusion is strongly supported by the observation (Section 3.1) that
many real correlations may be masked by noise due to too short signals
considered in practical applications.

\end{document}